**Simulation of Ion Irradiation of Nuclear Materials and Comparison with Experiment**

Z. Insepov [a], A. Kuksin [*], J. Rest [a], S. Starikov [a, b], A. Yanilkin [a, b], A. M. Yacout [a], B. Ye [a, c], D. Yun [a, c],

[a] Argonne National Laboratory, 9700 S. Cass Ave, Lemont, IL, 60439
[b] Joint Institute for High Temperatures RAS, 13 Izhorskaya , Moscow, 125412
[c] University of Illinois at Urbana-Champaign, 104 S. Wright Street, Urbana, IL 61801

**Abstract:** Radiation defects generated in various nuclear materials such as Mo and $CeO_2$, used as a surrogate material for $UO_2$, formed by sub-MeV Xe and Kr ion implantations were studied via TRIM and MD codes. Calculated results were compared with defect distributions in $CeO_2$ crystals obtained from experiments by implantation of these ions at the doses of $1\times10^{17}$ ions/cm$^2$ at several temperatures. A combination of in situ TEM (Transmission Electron Microscopy) and ex situ TEM experiments on Mo were used to study the evolution of defect clusters during implantation of Xe and Kr ions at energies of 150-700 keV, depending on the experimental conditions. The simulation and irradiation were performed on thin film single crystal materials. The formation of defects, dislocations, and solid-state precipitates were studied by simulation and compared to experiment. Void and bubble formation rates are estimated based on a new mesoscale approach that combines experiment with the kinetic models validated by atomistic and Ab-initio simulations. Various sets of quantitative experimental results were obtained to characterize the dose and temperature effects of irradiation. These experimental results include size distributions of dislocation loops, voids and gas bubble structures created by irradiation.

---

[*] Permanent address: Joint Institute for High Temperatures RAS, 13 Izhorskaya , Moscow, 125412

# 1. INTRODUCTION

Recently, a new application of ion implantation has been initiated that explores a unique capability in studying radiation effects and explore kinetics of defects in nuclear reactor environments [2]. This is a new and important research field since sustainable nuclear energy production must include a comprehensive analysis of nuclear fuel behavior. Fuel behavior codes are sensitive to materials parameters, many of which have large uncertainties, or have not been measured and, thus, a complete understanding of radiation damage and swelling of nuclear fuels throughout the operating burnup and temperature regime is required [3].

Atomistic simulation is a powerful method for tracking defect accumulation during ion implantation, and for estimating the values of critical materials properties and parameters used in kinetic fuel-behavior models. Whereas first principle simulations are limited to a few hundred atoms at most, classical molecular dynamic (MD) calculations with many millions of atoms are routinely performed. However, the reliability and predictive power of classical MD depend crucially on the quality of the effective potential employed. In the case of elementary solids, such potentials are usually obtained by adjusting a few potential parameters to optimally reproduce a set of reference data, which typically includes a number of experimental values such as lattice constants, cohesive energies, or elastic constants, sometimes supplemented with ab-initio cohesive energies and stresses. However, in the case of more complex systems with a large variety of local environments and many potential parameters to be determined, such an approach cannot help. Here a new method of deriving realistic interatomic potentials is presented.

# 2. SIMULATION

The force matching method (FMM) provides a way to construct physically justified potentials even under such circumstances as absence of experimental data. This method provides new interatomic potentials which are obtained on the basis of synthesis of quantum and statistical mechanics. The idea is to compute force and energies from first principles for a suitable selection of small reference system and to adjust the parameters of the interatomic potential to optimally reproduce them [4, 5]. The method allows creating correct potentials for simulation of various processes such as phase transitions, deformation at different temperature, and fracturing.

With the use of FMM-procedure, we developed an interatomic potential for Mo-Xe systems. The reference data was calculated by VASP code [6]. The following parameters were used: the electron orbitals were represented using plane-waves, with a cut-off energy of 400 eV; the generalized gradient approximation (GGA) for pseudopotential; 2x2x2-points in k-space. We used 81 various configurations with total number of atoms equal 10746. These configurations concluded a different Mo-Xe systems: 39 states with pure Mo (liquid and bcc solid states at different densities; solid states with SIAs and/or vacancies and/or surface), 20 states with pure Xe (liquid and solid states) and 22 states with Mo-Xe (including a single Xe atom in pure Mo). Generally the matching was carried out with three value types: energy (only one in every configuration); stress tensor (6 in every configuration); and forces (fx, fy and fz per every atom in configurations). Only the energy at matching was used in the configurations with SIA and vacancies in Mo. Table 1 shows comparison of the Mo parameters calculated using the new many-body Mo potential with experimental data.

Subsequently, the force matching procedure was implemented. The potential is realized in EAM (Embedding Atom Method) form with seven independent functions. In our investigation the potential functions were set by splines with 10 independent parameters. The search algorithm for potential parameters was obtained with the *potfit*-code [5]. A search was performed for a minimum of the target function

$$Z = \sum_i^N \sum_{\alpha=x,y,z} \frac{(f_{i\alpha} - f_{0,i\alpha})^2}{f_{0,ix}^2 + f_{0,ix}^2 + f_{0,ix}^2 + \varepsilon}, \tag{1}$$

where $f_{0,\alpha i}$ and $f_{\alpha i}$ are value of force components of atom $i$ calculated by VASP. Summation is carried out over atoms and configurations in the reference data.

The derived potential reproduces the reference data with good precision. For instance, energy difference equals 0.01 eV (precision about 0.1 %). The new interatomic potential simulates experimental equations of states of pure Mo and pure Xe. A comparison of simulation results with the new potential with some experimental data were performed for verification. Table 2 shows comparison of the defect properties calculated with the new EAM-potential with the VASP data.

Note that a difference between formation energies of defects is more important than total energies. Two configurations of SIA with minimal energy (Crowdion and Dumbbell <111>) have almost equal values. This fact explains a high one-dimensional mobility of SIA in pure Mo at low temperatures. In addition the new interatomic potential reproduces experimental results on scattering Mo atoms by Xe atoms (see figure 1).

Figure 1 compares the sputtering yields for Xe+ ion bombardment of a Mo (100) surface calculated in the present paper with the experimental data obtained from the literature [7–12]. We used three relatively new interatomic potentials for description of the Mo-Xe interaction. The parameters for the pair potential function for set #1 yield data close to experiment at higher energies, namely, $E$~100 eV. However, the calculated yields are much higher than the data at energies lower than 60 eV. Set #2 gives calculated yields close to experiment for both high and low energy regions. In addition, a new many-body interatomic potential for Mo-Xe system was also developed that gave close sputtering yields compared with experimental data in Fig. 1.

In what follows, the set #2 based on pair potential and the new many-body Mo-Xe potentials were applied for studying Xe-bubble properties in Mo. The Mo-Mo embedded-atom method (EAM) potential presented in this work reproduces the cold curve in agreement with the experimental data up to approximately 600 GPa (corresponding compression $V/V_0$ ~ 0.5). In addition, the description of thermal expansion is replicated well up to the melting point. The most stable configurations of interstitial defects in Mo are <111> dumbbell and <111> crowdion with very small differences in formation energies. This configuration provides for one-dimensional migration of self-interstitial atoms at very low temperatures [13] in agreement with the resistivity recovery measurements following electron irradiation [14]. With increasing temperature the <111>-<110> dumbbell transitions are activated, providing a rotation of the axis of migrating crowdions, and hence providing a basis for a viable mechanism for three-dimensional diffusion.

Fast parallel calculations were carried out on a Blue Gene supercomputer by using a Lammps MD simulation package [15]. A typical MD simulation of a system containing 22,000 Mo atoms in 10 ns was completed in six hours of computing time.

Fig. 2 shows the distribution of vacancies and self-interstitial defects along the ion track in pure Mo irradiated with an Xe+ ion, with energy of about 40 keV. The volume of the basic simulation cell was chosen to be 20×20×60 nm$^3$.

In order to estimate the diffusivity an initial point defect (vacancy or self interstitial atom) have been created within the simulation box of size from 10x10x10 to 30x30x30 lattice constants containing defect-free *bcc* crystal under periodic boundary conditions. Diffusion coefficients of atoms (tracers diffusivity) due to defects are calculated from the coordinates of the particles in the system. The diffusivity of atoms due to vacancies or interstitials is physically equivalent to the tracer diffusion coefficient in a pure crystal and is connected to the diffusivity of defects themselves $D_\alpha$ ($\alpha$ = v, i denoting vacancy or SIA) by the correlation factor *f*: $D_t = f D_\alpha$ (see [16] for details). The correlation factor can be evaluated theoretically for vacancies. It gives a possibility to calculate diffusion coefficient for vacancies.

The diffusivity mechanism in case of intestitials is complex and the analysis if the trajectory of the defects is required. The position of SIA *R(t)* is determined as the position of atom with the highest potential energy and taking into account the periodic boundary conditions. The diffusion coefficient is calculated in accordance with the formula: $D_i = \lim_{t \to \infty} \langle \overline{R}^2(t) \rangle / 6t$. Averaging is performed over an ensemble of realizations of the migration of SIA: a long trajectory (~ 10 ns) is divided into a parts, which are considered as an independent realizations.

The data at low temperatures can be described by an Arrhenius formula, corresponding activation energy is 0.02 eV for SIA with a new Argonne potential [17]. One can see that diffusion is activated at very low temperatures for EAM potentials by Starikov and Dudarev et al. in accordance with the experimental observations [14], where the temperature for the onset of long-range migration for Mo was determined as 35 K.

Figs. 3a-c show the diffusion coefficients of vacancies, self-interstitials and finally the self-diffusion coefficient calculated by MD in this work by using the expression [16]

$$D_{self} = f_v D_v + f_{sia} D_{sia},$$

and the result was compared with two existing experimental data sets [17, 18].

Fig. 4 shows the collision cascade showing the most of the track area is build up with vacancies and the self-interstitial clusters are formed in the outer regions.

### 3. EXPERIMENTAL

Ion irradiation was done in the IVEM-Tandem facility at Argonne National Laboratory, and the scanning transmission electron microscopy (STEM) was carried out on JEOL 2200Fs TEM at the University of Illinois. 500 keV Xe ions were implanted into single crystal $CeO_2$ TEM specimen at 600°C to an accumulated dose of $2 \times 10^{16}$ ions/cm$^2$. Fig. 5 shows the atomic level crystal structures of the specimen before and after irradiation. The specimen was in a perfect crystal structure before irradiation (Fig.5 (a)) and remains crystalline after bombardment as displayed in Fig. 5 (b). The areas with darker contrast in Fig.5 (b) suggest formation of defect clusters, which are in a size range of 1 – 3 nm in diameter. However, the nature of these defect clusters is not clear yet. Therefore, any direct comparison of these TEM experiments to MD simulations is not yet possible since these two methods provide information on very different time scales.

Since the irradiation dose in simulation was much lower ($1\times10^{12}$ ions/cm$^2$) the sizes of the clusters are smaller. The detailed analysis shows that they can be formed by two or three interstitial dislocation loops with the diameters of 10-15 Å.

Fig. 6 and 7 show Xe gas bubble size distributions obtained at different ion irradiation dose levels on 5% La doped CeO$_2$ single crystal thin film and 25% La doped CeO$_2$ single crystal thin film respectively. The thin films were grown on SrTiO$_3$ substrate with Molecular Beam Epitaxial (MBE) technique. The Xe ion implantations were carried out with 700keV *ex situ* irradiations and 500keV *in situ* irradiations at 600C. The sizes of Xe gas bubbles were measured at implantation depths consistently around 80-100nm on planar view La doped CeO$_2$ single crystal thin film specimens after Xe ion implantations. The effective diameters of gas bubbles were obtained by measuring the area of the gas bubble features with image processing software ImageJ. Five sets of 32.5nm by 32.5nm boxes were drawn and number densities of Xe gas bubbles were measured with bubble feature profiling by ImageJ as well. The error bars in the above shown figures are representations of statistical errors within the five independent measurements.

These experimental bubble distribution results are crucial to benchmark computer simulation results such as those from kinetic theory models or kinetic Monte Carlo models proposed in this study.

## 4. SUMMARY

Kinetic mesoscale models, such as those developed at Argonne National Laboratory are directly comparable to reactor experiments. Our new concept is based on kinetic rate-equations for radiation damage, energetics and kinetics of defects, and swelling of fuels as a function of temperature and burnup. Quantum and classical atomistic simulation methods are applied to increase our understanding of radiation damage and defect formation and growth processes and to calculate the probabilities of elemental processes and reactions applicable to irradiated nuclear materials. Since the interaction potentials are critical for the new concept, they were developed based on a force-matching method data from ab initio calculations or were fitted to existing experimental data.

In the present paper, a new many-body potential is proposed for pure Molybdenum that consists of using ab-initio and atomistic MD simulation methods verified against existing surface erosion experimental data. Several new Xe-Mo potentials were also parameterized via comparison of the calculated sputtering yield of a Mo-surface bombarded with Xe ions with experimental data. Diffusion in irradiated Mo was studied for voids and Xe-bubbles and compared to unirradiated Mo.

**Acknowledgments**

This work was supported by the Office of Advanced Scientific Computing Research, Office of Science, under Contract DE-AC02-06CH11357.

Table I: Comparison of simulation results (with new potential) with experimental data for pure Mo

|  | Cohesive energy (eV) | Lattice parameters (Å) | C11 (GPa) | C12 (GPa) | Melting temperature (K) |
|---|---|---|---|---|---|
| Simulation | 6.91 | 3.1469 | 560 | 225 | 2630 |
| Experiment | 6.82 | 3.15 | 464 | 163 | 2890 |

Table II: Comparison of simulation results (with new potential) with VASP calculation for Mo defects

|  | Formation energy of Crowdion <111> (eV) | Formation energy of Dumbbell<111> (eV) | Formation energy of Dumbbell<110>(eV) | Formation energy of vacancy |
|---|---|---|---|---|
| Simulation | 6.42 | 6.43 | 6.67 | 2.79 |
| VASP | 6.89 | 6.88 | 7.02 | 2.40 |

**Figure captions:**

Figure 1. Comparison of the sputtering yield of a Mo (100) surface bombarded by accelerated Xe+-ions interacting with Mo atoms via a Morse potential with the experimental data from refs. [33-38] (see also Fig. 2).

Figure 2. Distribution of vacancies in Molybdenum calculated via Molecular Dynamics for a Xe+ ions, with energy of 25 keV.

Figure 3a-c. The diffusion coefficients of vacancies, interstitials and the self-diffusion coefficient of Molybdenum compared with two experiments. Fig. 3a) The diffusion coefficient of vacancies calculated in this work can be approximated by expression: $D_v = 2.44835e-2 \, \text{Exp}[-1.1322e4/T,K]$ in cm2/s; Fig. 3b) The self-interstitial diffusion coefficient of Molybdenum calculated in this work by MD, can be approximated at low temperatures : $D_i = 7.18229e-4 \, \text{Exp}[-1.79861e2/T,K]$, cm2/s.

Figure 4. MD simulation of collision cascade and radiation damage of single crystal Mo: (a) Initial track formation at time instant of 0.05 ps; (b) 0.5 ps after irradiated with 50 keV Xe to a dose of $1 \times 10^{12}$ ions/cm$^2$ at 300°C; (c) Clusters of interstitial atoms formed by association of SIAs within 3 nanoseconds of simulation.

Figure 5. Scanning TEM micrograph of single crystal $CeO_2$ (a) before irradiation (b) after irradiated with 500 keV Xe to a dose of $2 \times 10^{16}$ ions/cm$^2$ at 600°C. Electron beam direction is along <001>.

Figure 6. Bubble size distribution from 700keV and 500keV Xe irradiations at 600C on 5% La doped $CeO_2$ (lines are drawn between data points to only guide the eyes)

Figure 7. Comparison of bubble size distributions between different doses with 700keV Xe *ex situ* irradiations and 500keV Xe *in situ* irradiations at 600C on 25% La doped $CeO_2$ (lines are drawn between data points to only guide the eyes)

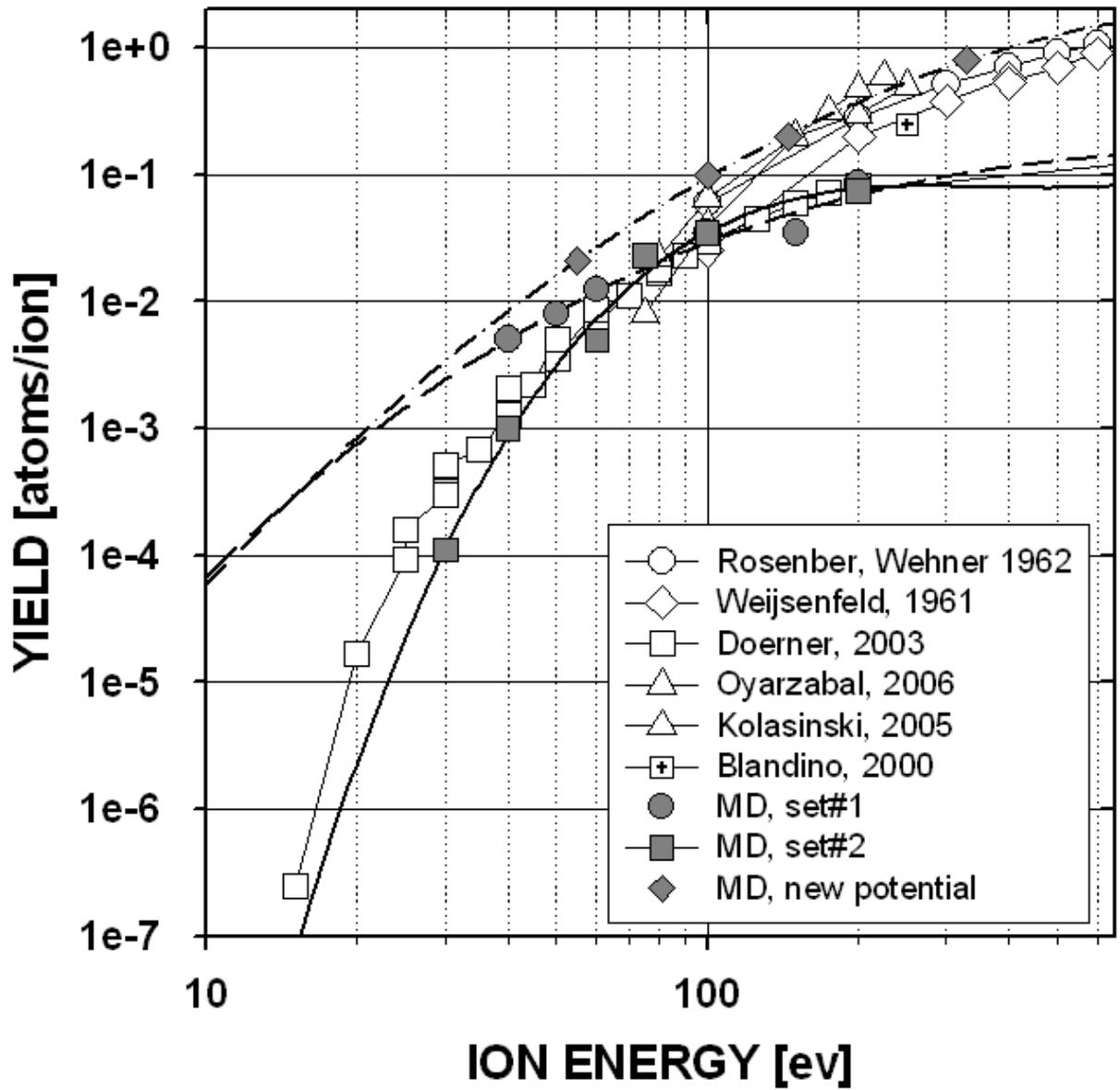

Figure 1

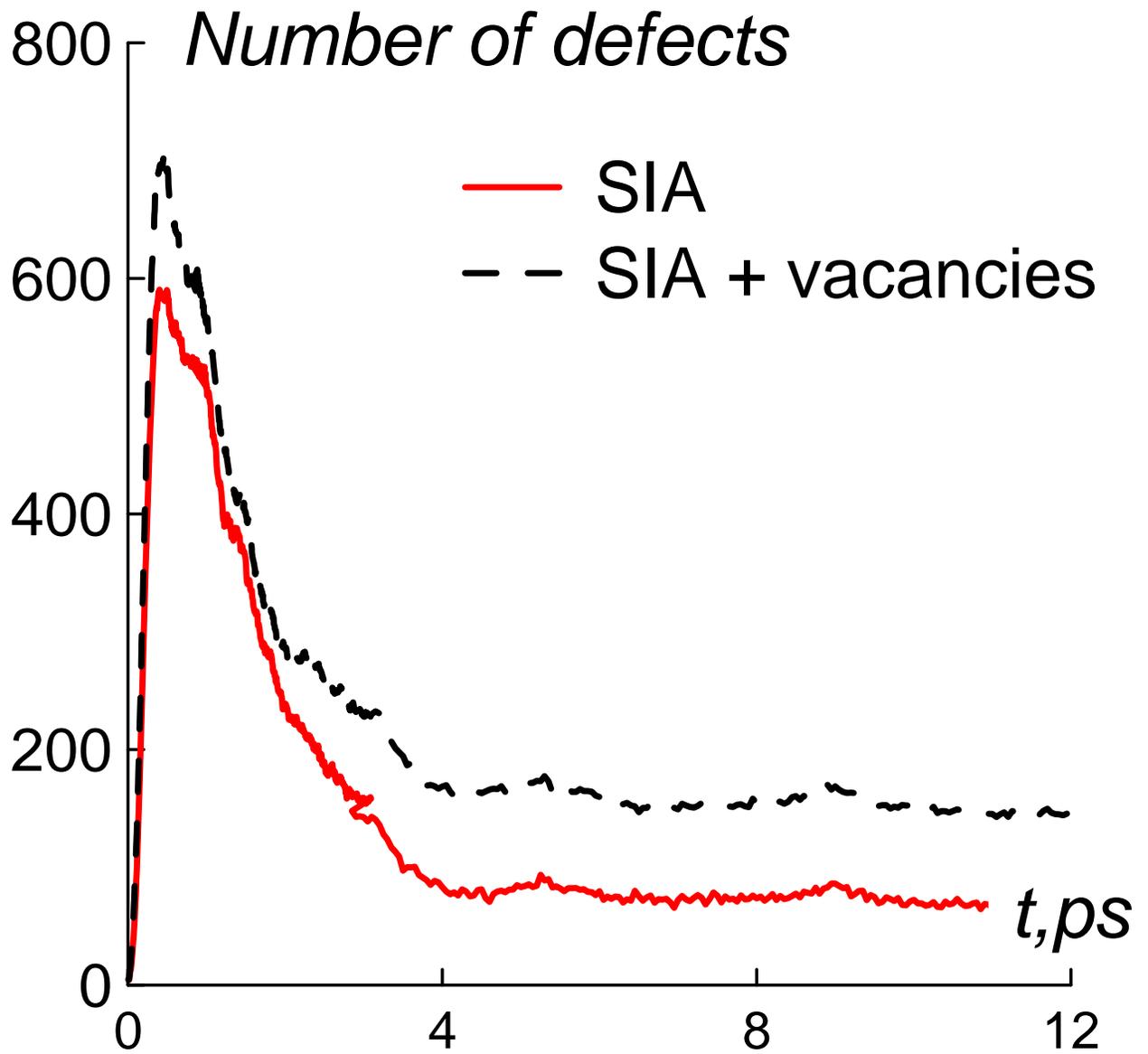

Figure 2

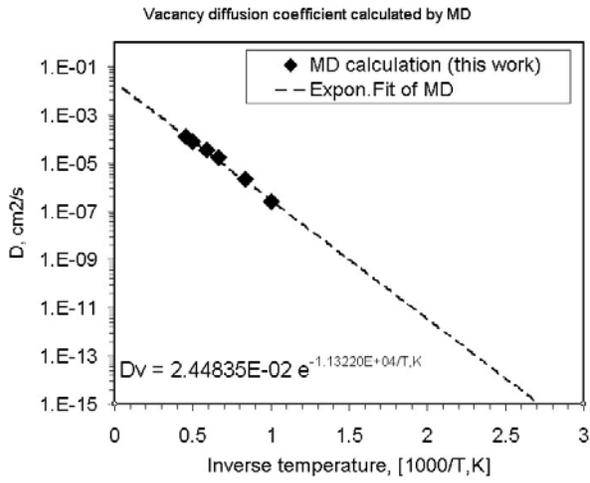 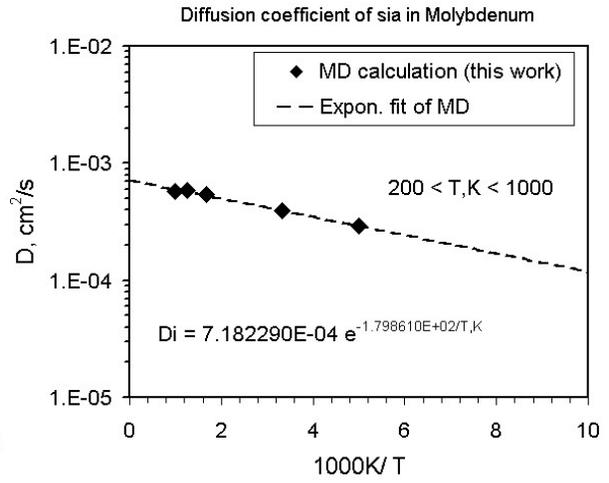

Figure 3a                                       Fig. 3b.

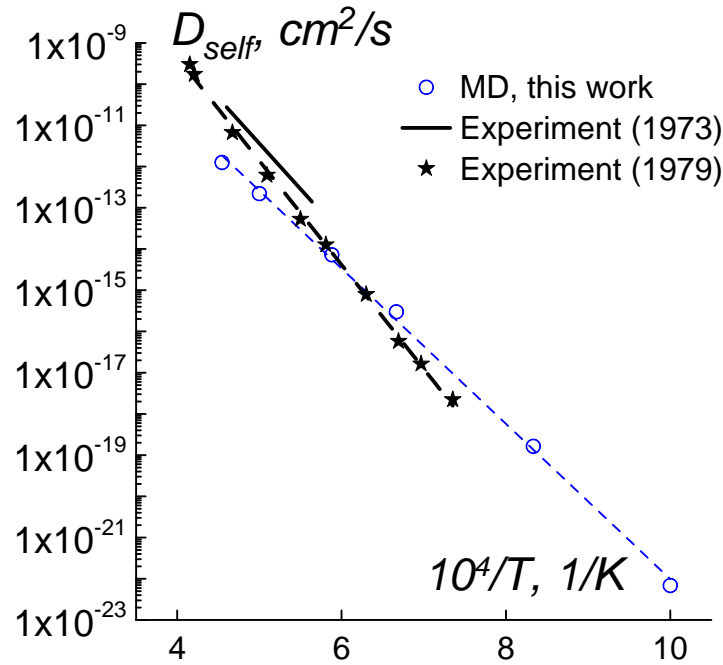

Figure 3c

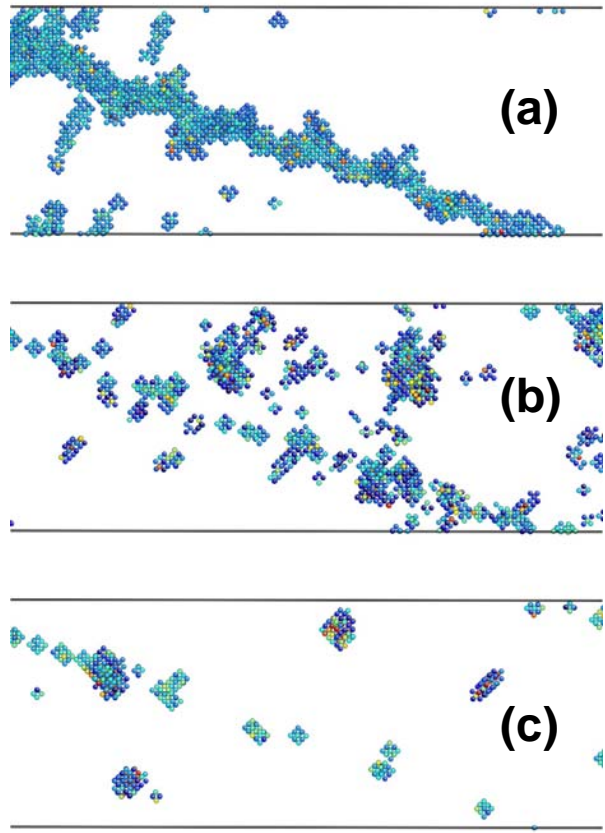

Figure 4

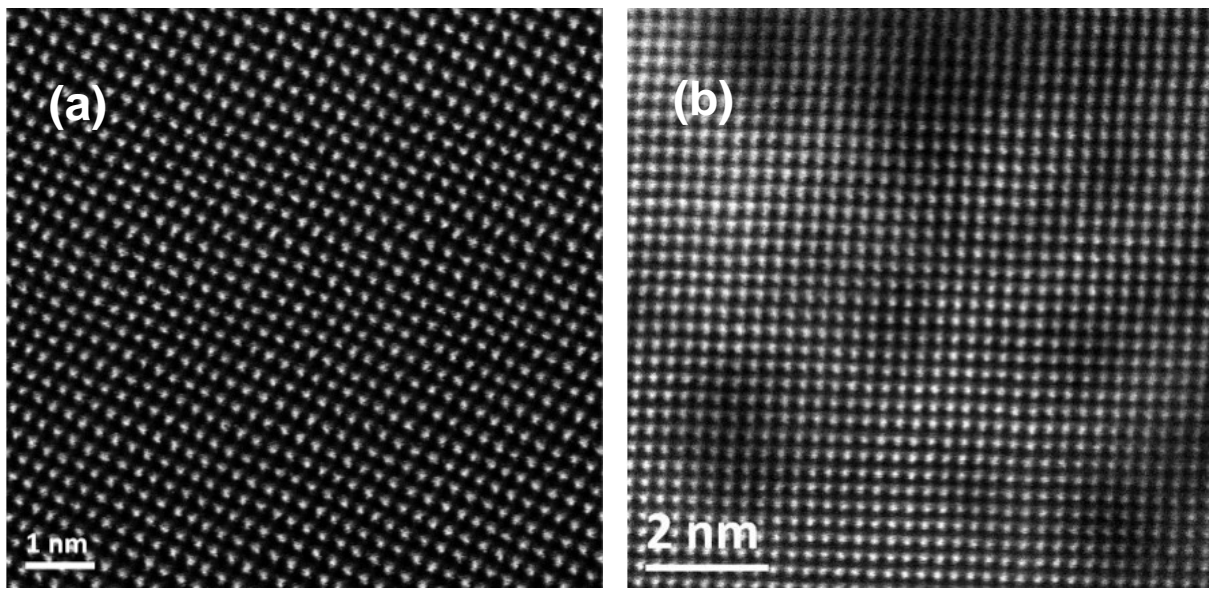

Figure 5

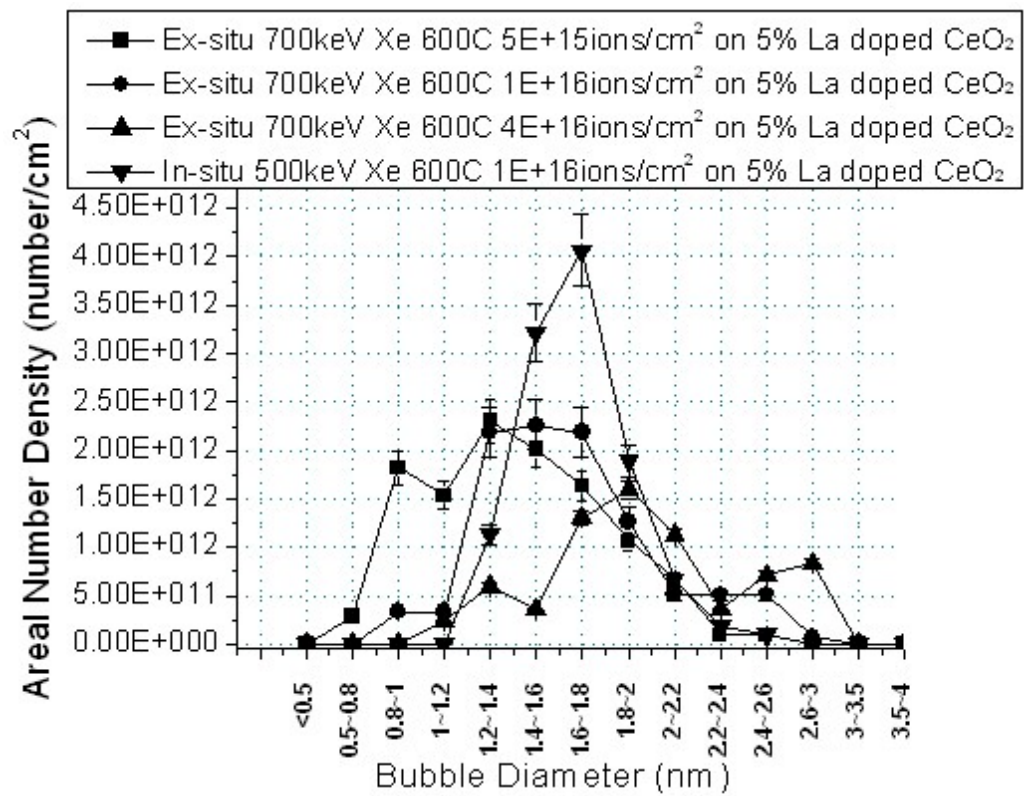

Figure 6.

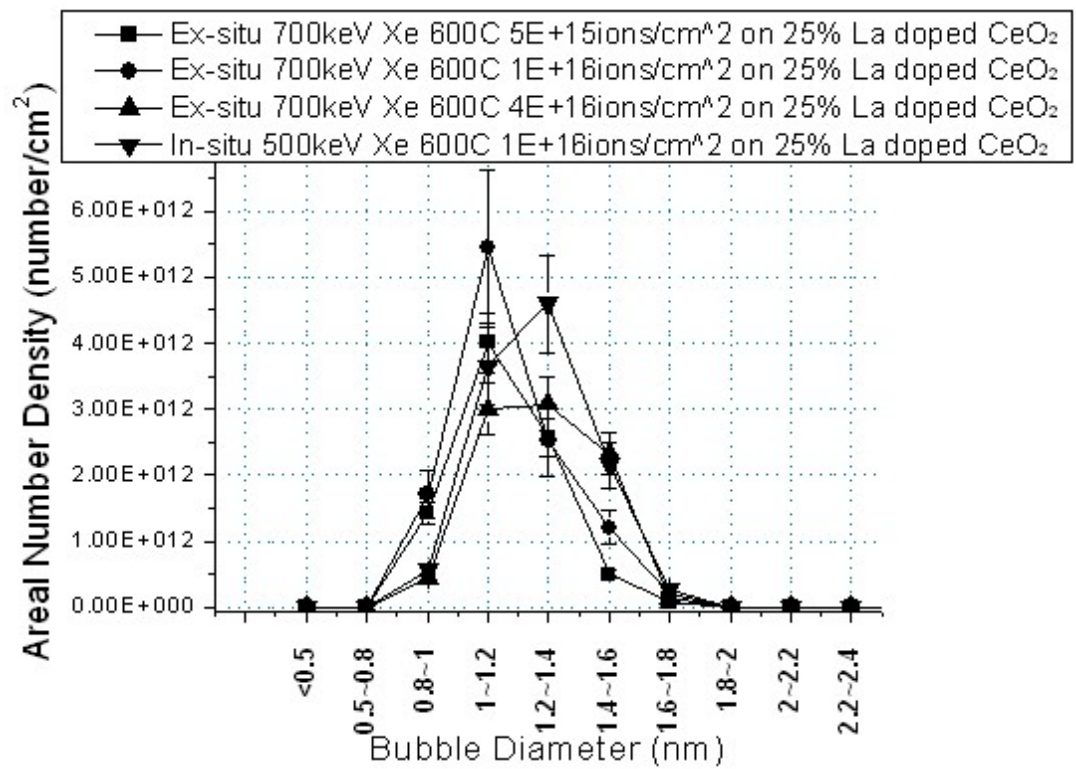

Figure 7.